\documentclass[fleqn,usenatbib]{mnras}

\usepackage{newtxtext,newtxmath}

\usepackage[T1]{fontenc}
\usepackage{ae,aecompl}

\usepackage{graphicx}	
\usepackage{amsmath}	
\usepackage{amssymb}	

\title[MW Progenitors]{Reverse engineering the Milky Way}

\author[D. A. Forbes]{
Duncan A. Forbes,$^{1}$\thanks{E-mail: dforbes@swin.edu.au}
\\
$^{1}$Centre for Astrophysics \& Supercomputing, Swinburne University, Hawthorn, VIC 3122, Australia
}

\date{Accepted XXX. Received YYY; in original form ZZZ}

\pubyear{2019}

\begin{document}
\label{firstpage}
\pagerange{\pageref{firstpage}--\pageref{lastpage}}
\maketitle

\begin{abstract}

The ages, metallicities, alpha-elements and integrals of motion of globular clusters (GCs) accreted by the Milky Way from disrupted satellites remain largely unchanged over time. Here we have used these conserved properties in combination to assign 76 GCs to 5 progenitor satellite galaxies -- one of which we dub the Koala dwarf galaxy. We fit a leaky-box chemical enrichment model to the age-metallicity distribution of GCs, deriving the effective yield and the formation epoch of each satellite. Based on scaling relations of GC counts we estimate the original halo mass, stellar mass and mean metallicity of each satellite. The total stellar mass of the 5 accreted satellites contributed around 10$^{9}$ M$_{\odot}$ in stars to the growth of the Milky Way but over 50\% of the Milky Way's GC system. The 5 satellites formed at very early times and were likely accreted 8--11 Gyr ago, indicating rapid growth for the Milky Way in its early evolution. 
We suggest that at least 3 satellites were originally nucleated, with the remnant nucleus now a GC of the Milky Way. 
Eleven GCs are also identified as having formed ex-situ but could not be assigned to a single progenitor satellite.

\end{abstract}




\section{Introduction}

It has long been recognised that the accretion of satellite galaxies has contributed to the growth of the  Milky Way (Searle \& Zinn 1978). In a recent review, Bland-Hawthorn \& Gerhard (2016) concluded that the Milky Way (MW) had accreted around one hundred satellite galaxies, with the bulk of the mass growth due to a few large satellites accreted at early times. 
The contribution of globular clusters (GCs) to the Milky Way GC system is also dominated by the more massive satellites as galaxies with stellar masses of less than $\sim$10$^7$ M$_{\odot}$ 
have few, if any, GCs (Forbes et al. 2018). Bland-Hawthorn \& Gerhard (2016) estimated that substructure (e.g. from disrupted satellites) within the Milky Way's halo accounts for some 45\% of the total halo stellar mass. Cosmological hydrodynamic simulations suggest that 15--40\% of the stars in a 10$^{12}$ M$_{\odot}$ halo (i.e. MW-like) were formed ex-situ and later accreted (Pillipich et al. 2018; Remus et al. 2020, in prep.). 

Given the potentially large contribution from infalling satellites to the growth of the MW galaxy, and to understand the relative role of low mass vs massive satellites, ideally one needs to `reverse engineer' the Milky Way's assembly history. 
Quantifying the properties of long ago accreted galaxies is problematic since  
after a few Gyr the disrupted satellites are no longer coherent in physical space.  However, their energies and angular momenta, or Integrals of Motion (IOM), are conserved, remaining relatively intact over time. For example, in the  simulation of Helmi \& de Zeeuw (2000)  satellites merging with the MW were still distinguishable in IOM space after 12 Gyr. The GCs from a 
given disrupted satellite will also clump together in IOM space. However, there can be significant overlap from one progenitor satellite to another and the boundaries are not known a priori. 

An additional signature of accreted GCs (which is conserved over time) is that they reveal a distinctive pattern in age and metallicity space. This age-metallicity relation (AMR) closely follows a leaky-box chemical enrichment model that reflects the enrichment of the original galaxy over time. Ages and metallicities are available for a large number of MW GCs. 

In principle, the alpha-element ratio [$\alpha$/Fe] offers an additional avenue to identify accreted GCs since they are expected to reveal lower [$\alpha$/Fe] ratios on average for a given [Fe/H] if formed in a low mass satellite galaxy (Hughes et al. 2019; Mackereth et al. 2019). Indeed, this seems to be the case 
(Pritzl et al. 2005; Recio-Blanco 2018) for MW GCs. The compilation by Recio-Blanco (2018), lists some 45 MW GCs with available [$\alpha$/Fe] ratios but in many cases only a few stars in each GC have been measured leading to relatively large uncertainties in their mean values. However, most recently Horta et al. (2020) have measured new alpha-element abundances for 46 GCs from the SDSS/APOGEE survey with 67 stars, on average, per GC. 

Thus, the combination of IOM and the AMR, and to a lesser extent alpha-element ratios, of GCs is a powerful one to identify disrupted satellites and hence reverse engineer the build-up of the MW galaxy. 

Previous work has identified accreted satellites from their GCs, based on their kinematics (e.g. Law \& Majewski 2010) but 
recently, the second data release from Gaia in 2018 has provided positions and velocities (i.e. all 6 dimensions) for a large sample of MW GCs. Massari et al. (2019; hereafter M19) have used this 
new information 
to assign GCs to a  progenitor satellite based on their location in IOM space. They also used, as a secondary criterion, the age-metallicity distribution of the GCs  to guide their selection.  

The M19 sample consists of 160 GCs, 151 of which have available kinematics, and for 69 they had ages and metallicities. Here we use their IOM analysis combined with the recent compilation of ages and metallicities from Kruijssen et al. (2019; hereafter K19), thus increasing the sample from 69 to 96 GCs. This increased list of GCs with ages and metallicities allows us to better constrain the GC membership of disrupted satellites and in some cases reassign the GC membership of M19. We are also guided by the alpha-element abundances measured by Horta et al. (2020) when available.

Here our approach is to start with the GCs identified 
as being formed in-situ within the early MW. Once these are removed from our sample, we then identify GCs associated with each satellite galaxy starting with the well-defined Sgr dwarf. Finally, we are left with a small number of GCs that may have come from several low mass satellites, each of which may have hosted as few as one GC. Using the number of GCs associated with each satellite galaxy we use scaling relations to estimate the halo mass, stellar mass and mean metallicity of the progenitor galaxy before it was accreted and hence reverse engineer the assembly of the Milky Way galaxy. 

\section{The Milky Way GC System} 

The Milky Way contains around 160 known GCs, although Minniti et al. (2019) have reported an additional 50 GC candidates in the direction of the Milky Way's bulge.
The recent compilation of K19 lists 96 GCs with ages and metallicities taken from three large studies. While the data and stellar population models applied are not homogeneous, the final list has  
typical uncertainties in age of $\pm$0.75 Gyr and $\pm$0.1 dex in [Fe/H]. 

\begin{figure}
	\includegraphics[width=0.8\columnwidth,angle=-90]{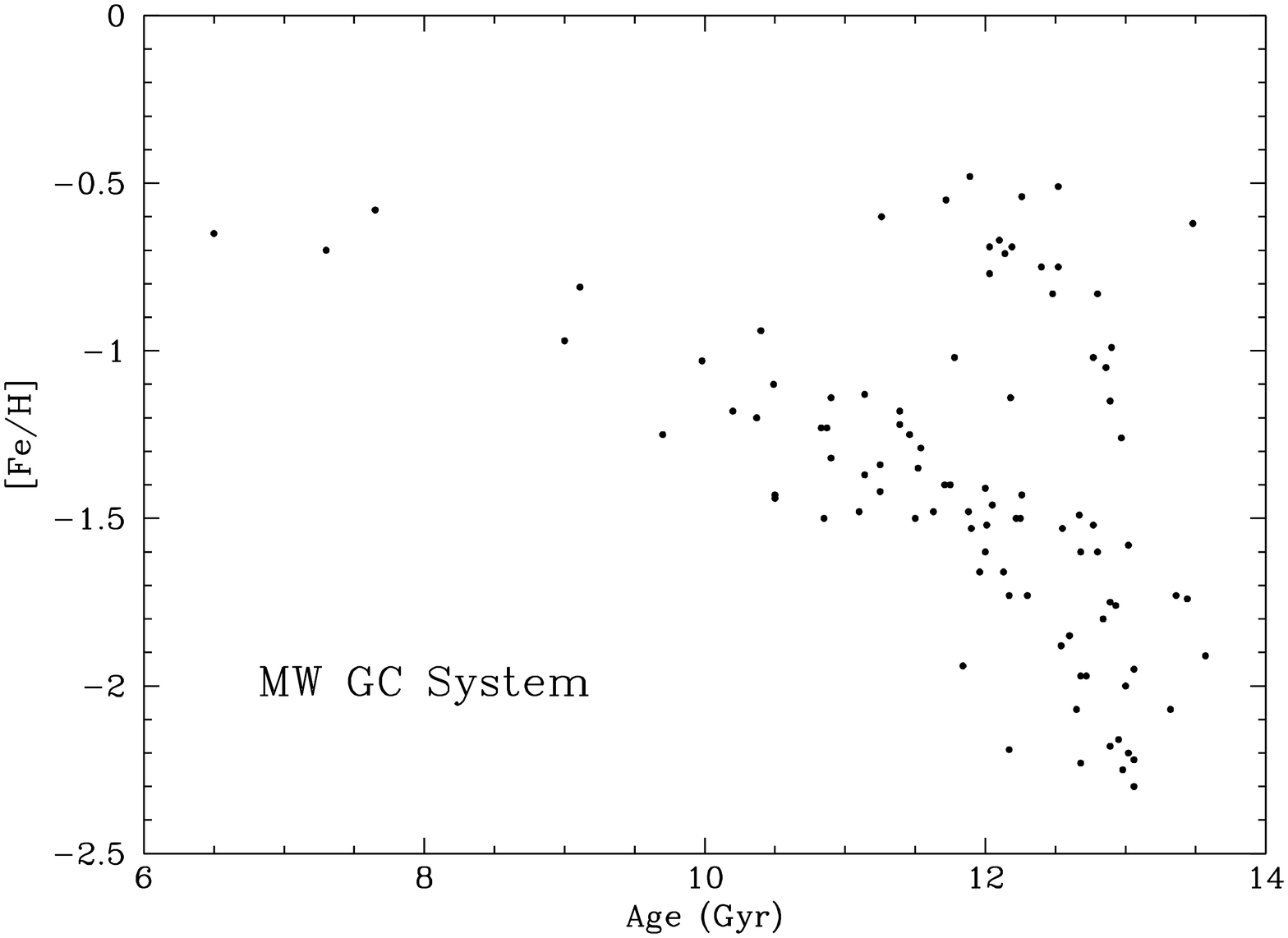}
    \caption{The age and metallicity distribution of the Milky Way globular cluster system. The main features are a near uniform old age group (associated with in-situ GCs) and a branch to younger ages/enhanced metallicities (associated with accreted GCs).   
    Typical uncertainties are $\pm$0.75 Gyr in age and $\pm$0.1 dex in [Fe/H]
     }
    \label{original}
\end{figure}

We show the ages and metallicities for these 96 GCs in Fig.~\ref{original}. This figure shows two well-known features, i.e. a near uniform old age group and a branch to younger ages and higher metallicities (e.g. Forbes \& Bridges 2010; Leaman et al. 2013a). 
The former are associated with in-situ GCs that enriched rapidly over a short period of time at early epochs in the main progenitor of the MW, and the latter are thought to have been accreted from disrupted dwarf satellites. Hereafter we refer to the main progenitor and  satellite branch, respectively. 

\section{Age-Metallicity Relation}

As shown in Forbes \& Bridges (2010), a simple chemical evolution model can provide a surprisingly good description of the 
age-metallicity relation of the GCs that were former members of an accreted satellite galaxy. Here we assume a leaky-box AMR similar to that used by M19 of the form:\\

[Fe/H] = $-p$ ~ln $(\frac{t}{t_{f}})$ \hspace{4cm}(1)\\

where $p$ is the effective yield of the system and 
$t _f$ is the look-back time when the system first formed from non-enriched gas. Below we fit an AMR in the form of eq. 1 to the GCs identified by M19 as being probable or tentative members of a disrupted satellite. We 
determine the best-fit
effective yield $p$ of the satellite (which tends to be larger in higher stellar mass galaxies) and the age of its initial formation. 


\section{In-situ Globular Clusters}

Of the 160 GCs listed by M19, 151 have Gaia kinematics. M19 define, based on their dynamical properties, a total of 62 GCs as formed within the main progenitor of the MW, i.e. in-situ. Of these, 26 are assigned to the in-situ disk and 36 to the in-situ bulge. 
The majority of these GCs do not have age or metallicity measurements available in K19. 

One GC (Pal~1), assigned by M19 to the main progenitor, lies on the satellite branch with age = 7.3 Gyr and [Fe/H] = --0.70. It is therefore very unlikely to be part of the main progenitor. We note that M19 did not have an age/metallicity for Pal~1 and that it lies close to the Gaia-Enceladus GCs in IOM space (below we re-assign it to Gaia-Enceladus).

On the other hand, three GCs have ages and metallicities that suggest they are also part of the main progenitor, i.e. E3 (age= 12.80 Gyr, [Fe/H] = --0.83), NGC 6121 (age = 12.18 Gyr, [Fe/H] = --1.14) and NGC 6441 (age = 11.26 Gyr, [Fe/H] = --0.60). M19 tentatively assigned E3 to the Helmi streams with NGC 6121 and 6441 assigned to the low-energy group. They are also close to the main progenitor in IOM space. Furthermore, Horta et al. (2020) find NGC 6121 and NGC 6441 (along with Pal~6) to have high alpha-element abundances and conclude that they formed in-situ (they did not have data for E3). Pal~6 does not have available age-metallicity data but its location in IOM space also overlaps with the main progenitor. We thus reassign all four GCs to the main progenitor.

\begin{figure}
	\includegraphics[width=0.8\columnwidth,angle=-90]{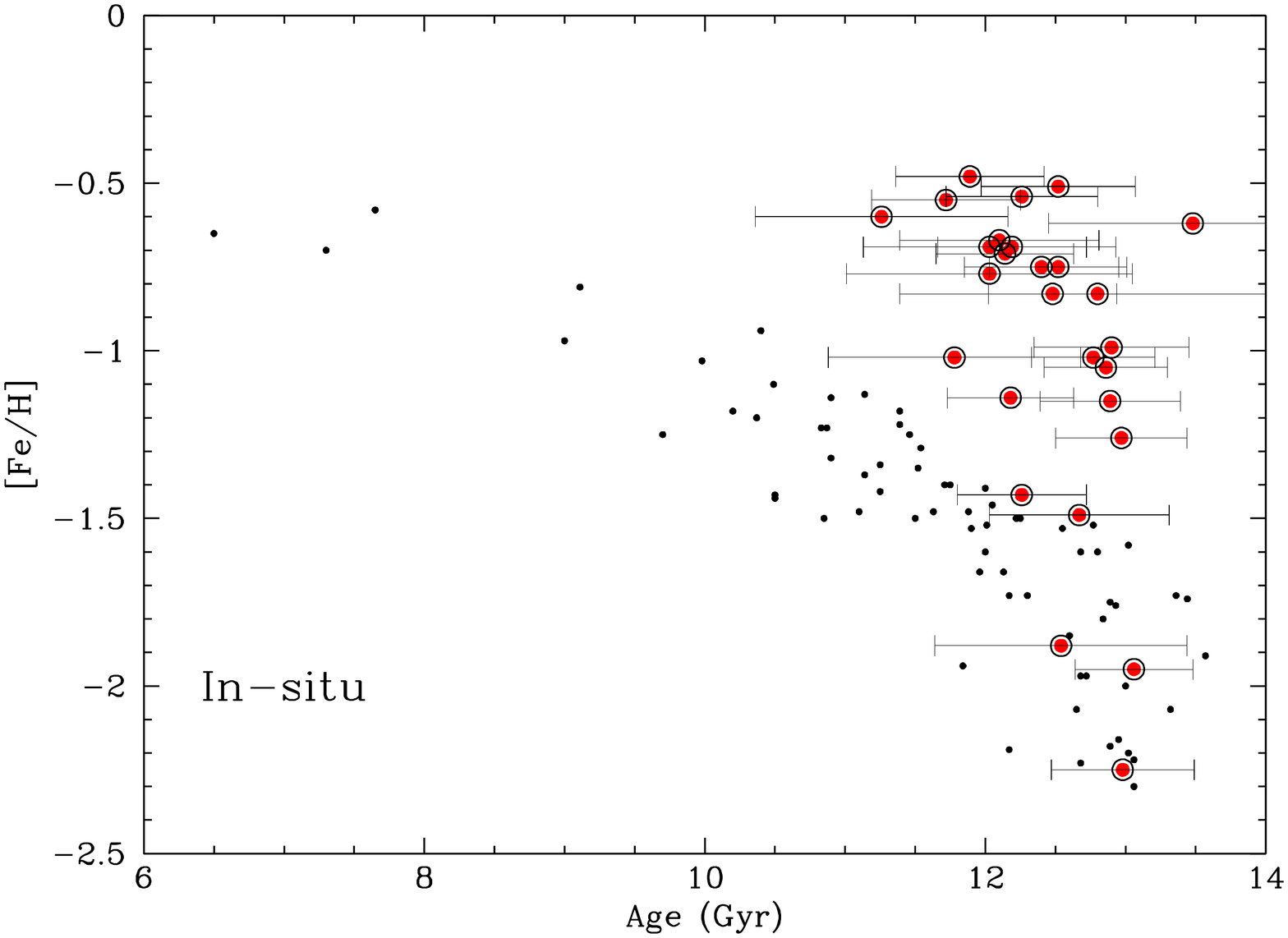}
    \caption{In-situ globular clusters. Red symbols show the GCs assigned to the main progenitor bulge and disk. 
     Small black symbols show all other GCs of the MW.}
    \label{main}
\end{figure}

From M19's initial total of 62 GCs we remove one and add four to give 65 in-situ formed GCs. We show the two dozen in-situ formed GCs with available ages and metallicities in Fig.~\ref{main}. 

We note that 8 GCs have no Gaia kinematic data nor age-metallicity data in K19 (one GC, AM4, has age-metallicity data but no Gaia kinematics). 
Typically these GCs are newly discovered and located in either the Galactic disk or bulge and thus are likely to be main progenitor in-situ GCs. 
If these 8 GCs are indeed in-situ formed, then a total of 73/160 (46\%) GCs were formed in-situ and 87/160 (54\%) were formed ex-situ and later accreted. 
These fractions lie between those found by M19 (i.e. 60\% accreted) but somewhat higher than those predicted by K19 (i.e. 43\% accreted). 

\section{Accreted GCs}

\subsection{The Sagittarius Dwarf}

Since its discovery by Ibata et al. (1994), the Sagittarius (Sgr) dwarf  galaxy has been associated with several GCs.
M19 find that 8 GCs are particularly well-defined in IOM space, i.e. 
NGC 2419, NGC 5824, Arp 2, Pal12, Terzan 7, Terzan 8, Whiting~1  
and the likely former nucleus M54 (NGC 6715). 

\begin{figure}
	\includegraphics[width=0.8\columnwidth,angle=-90]{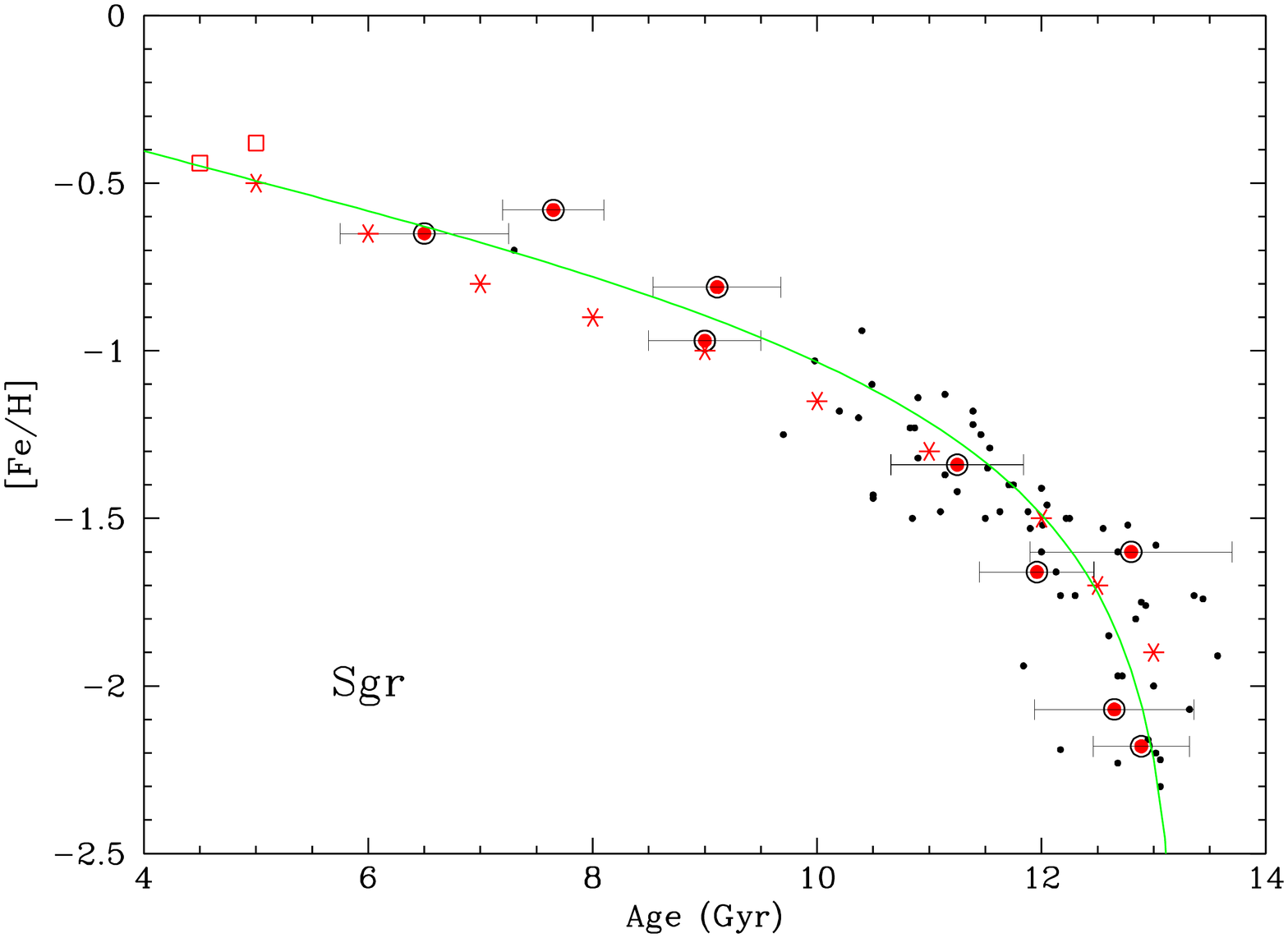}
    \caption{Sagittarius dwarf AMR. Red symbols show the 9 GCs we assign to the Sgr dwarf. The open squares correspond to two old open clusters and the asterisks to the field stars of Sgr from de Boer et al. (2015). Small black symbols show all remaining GCs of the MW after removing in-situ GCs. The green line is the best-fit AMR to the Sgr GCs. }
    \label{sgr}
\end{figure}

In Fig.~\ref{sgr} we show the ages and metallicities for these 8 Sgr GCs, along with two old open clusters (Berkeley 29 and Saurer 1) that appear to be associated with the Sgr dwarf in phase space (Carraro \& Bensby 2009) and the field star measurements of Sgr by de Boer et al. (2015).
Carraro \& Bensby (2009) suggested that the GC AM4 also belongs to the Sgr dwarf, on the basis that it agrees fairly well with the Law \& Majewski (2010) kinematic model. Gaia kinematics do not exist for AM4 and hence it was given no assignment by M19.
Its age (9.0 Gyr) and metallicity ([Fe/H] = --0.97) fits the general trend of the Sgr GCs.
Thus we fit an AMR to all 9 GCs using eq. 1 which gives an effective yield $p$ = 0.33 $\pm$ 0.06. This is very similar to the Sgr AMR found by FB10. 
The best-fit AMR for the GCs is also well matched to the two open clusters and follows the general trend of the field stars, albeit with a slight offset  (given the different stellar population models used we do not expect a perfect match between the stellar and GC ages and metallicities). The in-situ MW GCs  assigned in Section 4 have been excluded from Fig.~\ref{sgr}. In subsequent figures we additionally exclude all GCs that are assigned to a progenitor satellite. 

In summary, we assign the 8 GCs given by M19 above plus AM4 to the Sgr progenitor galaxy -- a dwarf spheroidal galaxy with M54 
(M$_V$ = --9.98, age = 11.25, [Fe/H] = --1.34) 
as its former nucleus. Infall into the Milky Way has been estimated to occur 
8 $\pm$ 1.5 Gyr ago by Dierickx \& Loeb (2017) and 9.3 $\pm$ 1.8 Gyr ago by Hughes et al. (2019). The stellar mass at infall has a range of 9.9--14.4 $\times$ 10$^7$ M$_{\odot}$ according to Niederste-Ostholt et al. (2012). The pre-infall halo mass is estimated by Gibbons et al. (2017) to be at least 6 $\times$ 10$^{10}$ M$_{\odot}$. 


\subsection{Gaia-Enceladus} 

\begin{figure}
	\includegraphics[width=0.8\columnwidth,angle=-90]{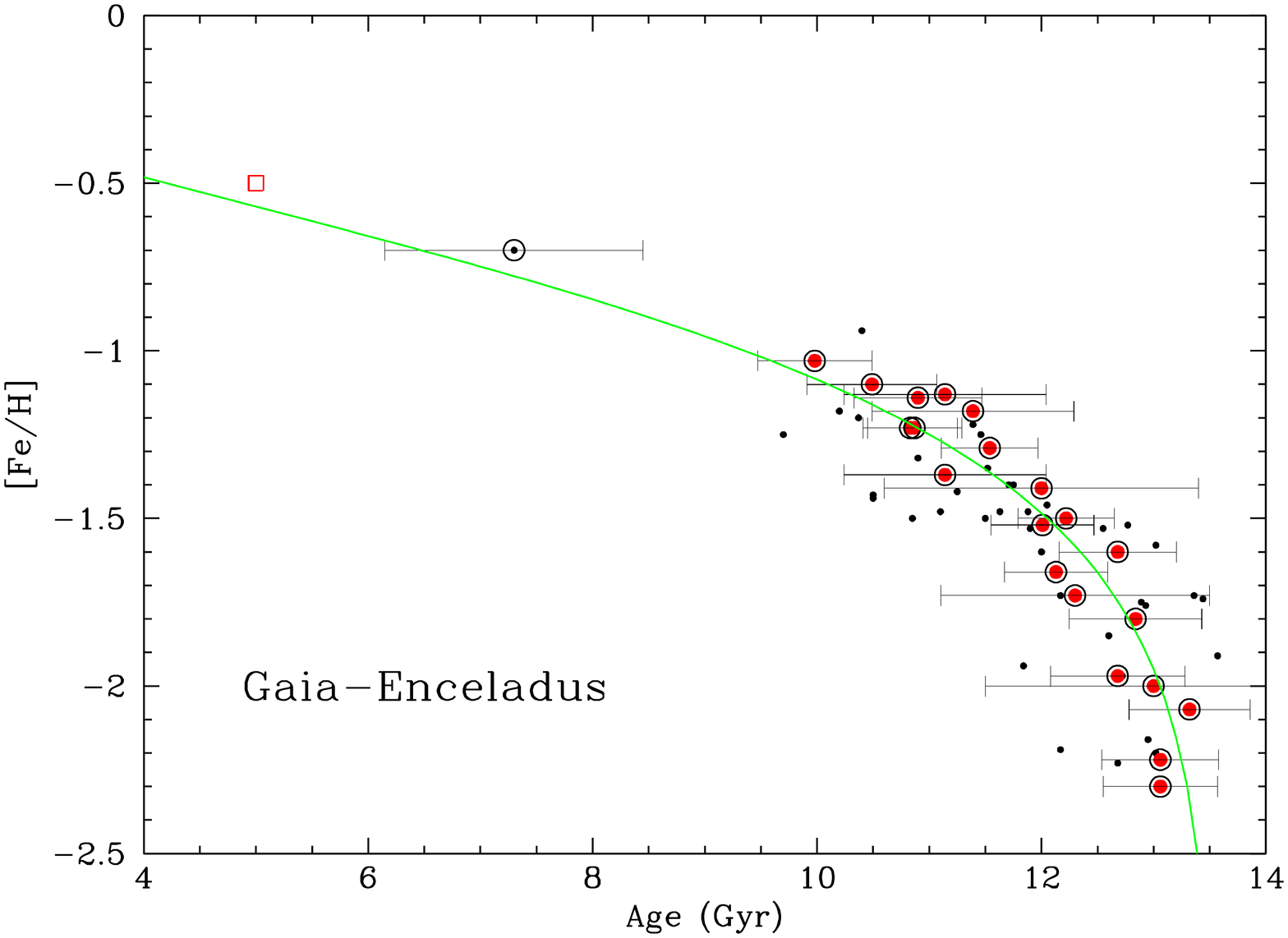}
    \caption{Gaia-Enceladus AMR. Red symbols show GC  assigned to Gaia-Enceladus, with the open symbol for the tentative GC Pal~1. Small black symbols show the remaining GCs of the MW. The red open square shows the old open cluster AM2. The green line is the best-fit AMR to the GCs. }
    \label{ge}
\end{figure}

One of the major results from  the Gaia satellite is the discovery of stellar debris (Helmi 2018; Myeong et al. 2018; Gallart et al. 2019) associated with a massive progenitor called the Gaia Sausage or Gaia-Enceladus. M19 associate 25 probable GCs,  2 tentative GCs and 5 dual GCs (i.e. possible membership of two host satellites). We display the 20 probable GCs with available age and metallicity data in Fig.~\ref{ge}. 
We are also able to include the tentative GC Pal 15 in Fig.~\ref{ge}, finding that it is consistent with the Gaia-Enceladus AMR. We thus add it to the list of probable GCs.
For the only other tentative GC, Pal 2, we have no age-metallicity information and continue to classify it as tentative.
We do not assign any of 5 dual member GCs to Gaia-Enceladus. M19 included $\omega$-Cen 
which is classed as either Gaia-Enceladus or Sequoia. As noted below, we associate $\omega$-Cen with Sequoia. Other dual membership GCs we associate with either Sequoia (i.e. NGC 3201, NGC 6101) or the Helmi streams (i.e. NGC 5904, NGC 5634) rather than Gaia-Enceladus, as explained below. We note that no age-metallicity estimate was available to M19 for NGC 5634 (age = 11.84 Gyr and [Fe/H] = --1.94).

Many of the GCs from M19 are in common with the prograde model for the Canis Major dwarf progenitor of Martin et al. (2004), i.e. NGC 1851, NGC 1904, NGC 2298, NGC 2808, NGC 5286, NGC 6205, NGC 6341, NGC 7078, NGC 7089 and IC 1257. Martin et al. also list Pal 1 which although listed by M19 as being part of the main progenitor, its age (7.3 Gyr) and metallicity ([Fe/H] = --0.70) make that very unlikely. 
The Gaia-Enceladus GCs are located close to the Main Progenitor in IOM space and hence the possible misclassification by M19 of Pal~1. Thus on the basis of the claim by Martin et al. (2004), its location in IOM space and its age and metallicity, we classify Pal~1 as a probable Gaia-Enceladus GC. 
This brings the total to 27 probable and 1 tentative GC. Our fit gives a well-defined AMR of yield $p$ = 0.27$\pm$0.02. 

Martin et al. (2004) also 
noted the possible connection with the old open cluster AM2. Fig.~\ref{ge} shows that this open cluster also fits the Gaia-Enceladus AMR extremely well. 

Although the most luminous GC associated with Gaia-Enceladus is NGC 2808 (M$_V$ = --9.4) and was suggested by Carballo-Bello et al. (2018) as the nucleus, we favour NGC 1851 (M$_V$ = -- 8.33, age = 10.49 Gyr, [Fe/H] = --1.10). NGC 1851 has peculiar stellar populations similar to those seen in M54 and $\omega$-Cen (Marino et al. 2014) and reveals 
an extended halo of stars (Olszewski et al. 2009; Kuzma et al. 2018). Bekki \& Yong (2012) showed that these properties can be successfully modelled with NGC 1851 being the remnant nucleus of a now stripped dwarf galaxy. In their simulation, the satellite galaxy was accreted at least 8.5 Gyr ago in order to strip away most of the galaxy stars leaving only a small halo around the remnant nucleus. Penarrubia et al. (2005) estimate the stellar mass at infall of Canis Major (which may be equivalent to Gaia-Enceladus) to be 6--20$\times$10$^8$ M$_{\odot}$. We note that Myeong et al. (2019) estimate a halo mass of 1--5 $\times$ 10$^{11}$ M$_{\odot}$, which is based on the mass of the GC system (similar to the approach we adopt in the next section).

\subsection{Sequoia}

Following M19, there is a fairly clear distinction in IOM space between Sequoia and Gaia-Enceladus. As noted by Myeong et al. (2019) Sequoia likely included GCs with retrograde orbits, including $\omega$-Cen which has been long suggested to be the former nucleus of an accreted dwarf galaxy (Freeman 1993). 
Here we accept the 5 GCs identified by M19 as probable Sequoia members (i.e NGC 5466, NGC 7006, IC 4499, FSR1758 and Pal 13). Age and metallicity measures are available for all except FSR1758 and Pal 13. M19 identify 4 tentative GC members, i.e. NGC 3201, NGC 6101, NGC 6535 and $\omega$-Cen (NGC 5139). 

\begin{figure}
	\includegraphics[width=0.8\columnwidth,angle=-90]{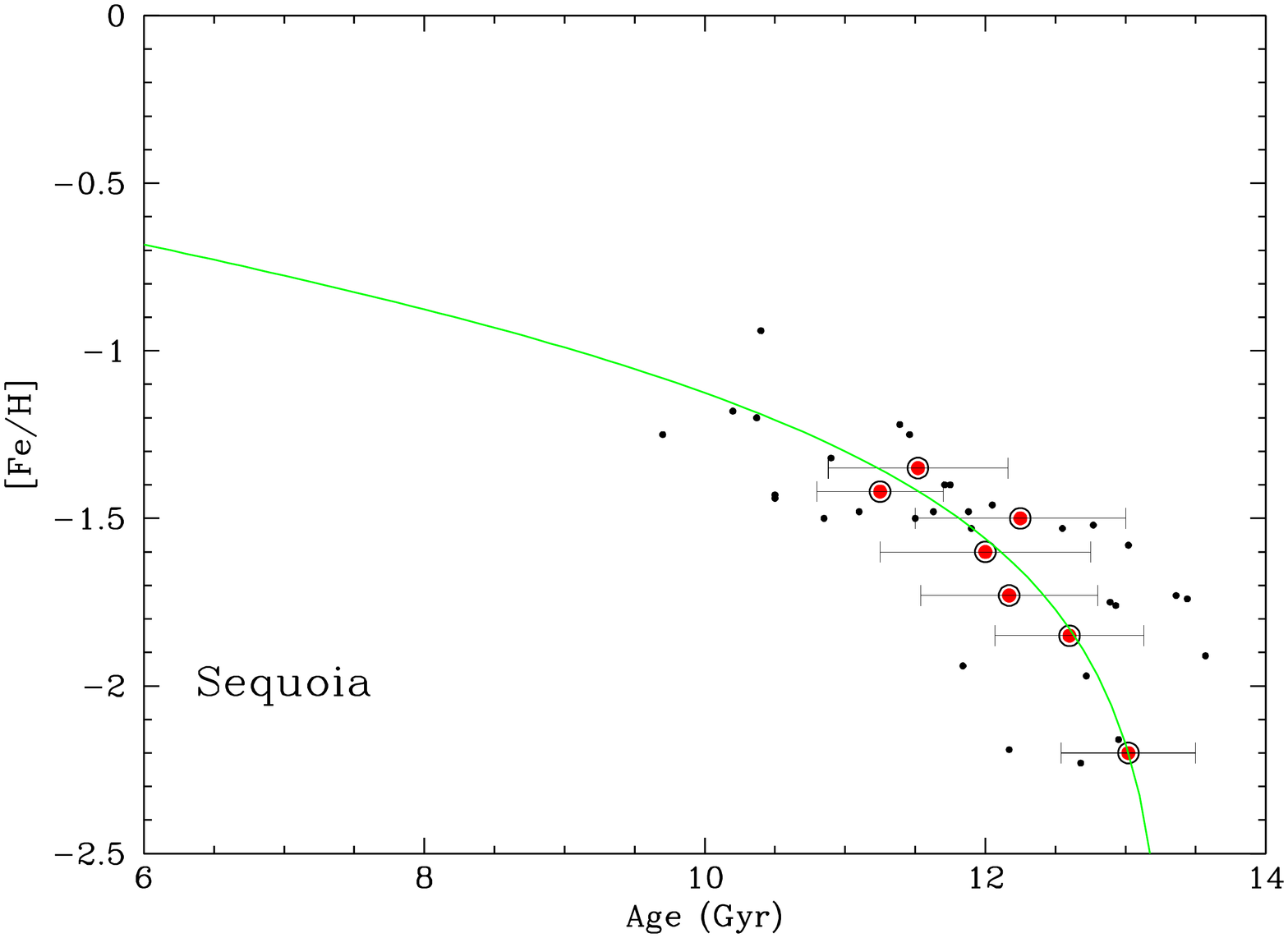}
    \caption{Sequoia dwarf AMR. Red symbols show GC assigned to Sequoia. Small black symbols show the remaining GCs of the MW. The green line is the best-fit AMR to the GCs. }
    \label{seq}
\end{figure}

In Fig.~\ref{seq} we show the AMR for the 3 probable GCs and 4 tentative GCs with age and metallicity measures available from K19. We find all 7 to be consistent with a a similar AMR with a yield of $p$ 
= 0.26$\pm$0.03 indicating a fairly low mass progenitor. Thus we change the assignment of the 4 tentative GCs by M19 to be probable GCs of Sequoia. As indicated in Table 1, we denote $\omega$-Cen (M$_V$ = --10.26, age = 11.52 Gyr, [Fe/H] = --1.35) as the former nucleus of the Sequoia progenitor galaxy. 

The infall time of Sequoia was around 10 Gyr ago from hydrodynamical modelling of $\omega$-Cen (Marcolini et al. 2007) and $\ge$ 9 Gyr according to Myeong et al. (2018). Both stellar population modelling (Valcarce \& Catelan 2011) and modelling of the associated S1 stellar stream (Myeong et al. 2018) both suggest a halo mass of $\sim$10$^{10}$ M$_{\odot}$. We note that Myeong et al. (2019) estimate a stellar mass of $\sim$5 $\times$ 10$^{7}$ M$_{\odot}$, which is based on the mass of the GC system (similar to the approach we adopt in the next section).


\subsection{Koala -- A Low Energy Progenitor}

M19 assigned 24 GCs with  particularly low energies and low perpendicular angular momenta to a single progenitor. 
As noted in section 4 we assign NGC 6121, NGC 6441 and Pal~6 to the main progenitor leaving 21 GCs.
In addition M19 classified NGC 6535 as either belonging to this low-energy group or Sequoia -- we have assigned this GC to Sequoia as mentioned above.  
For convenience, we name this low energy progenitor Koala (an Australian native animal known to sleep 18--22 hours each day). 

\begin{figure}
	\includegraphics[width=0.8\columnwidth,angle=-90]{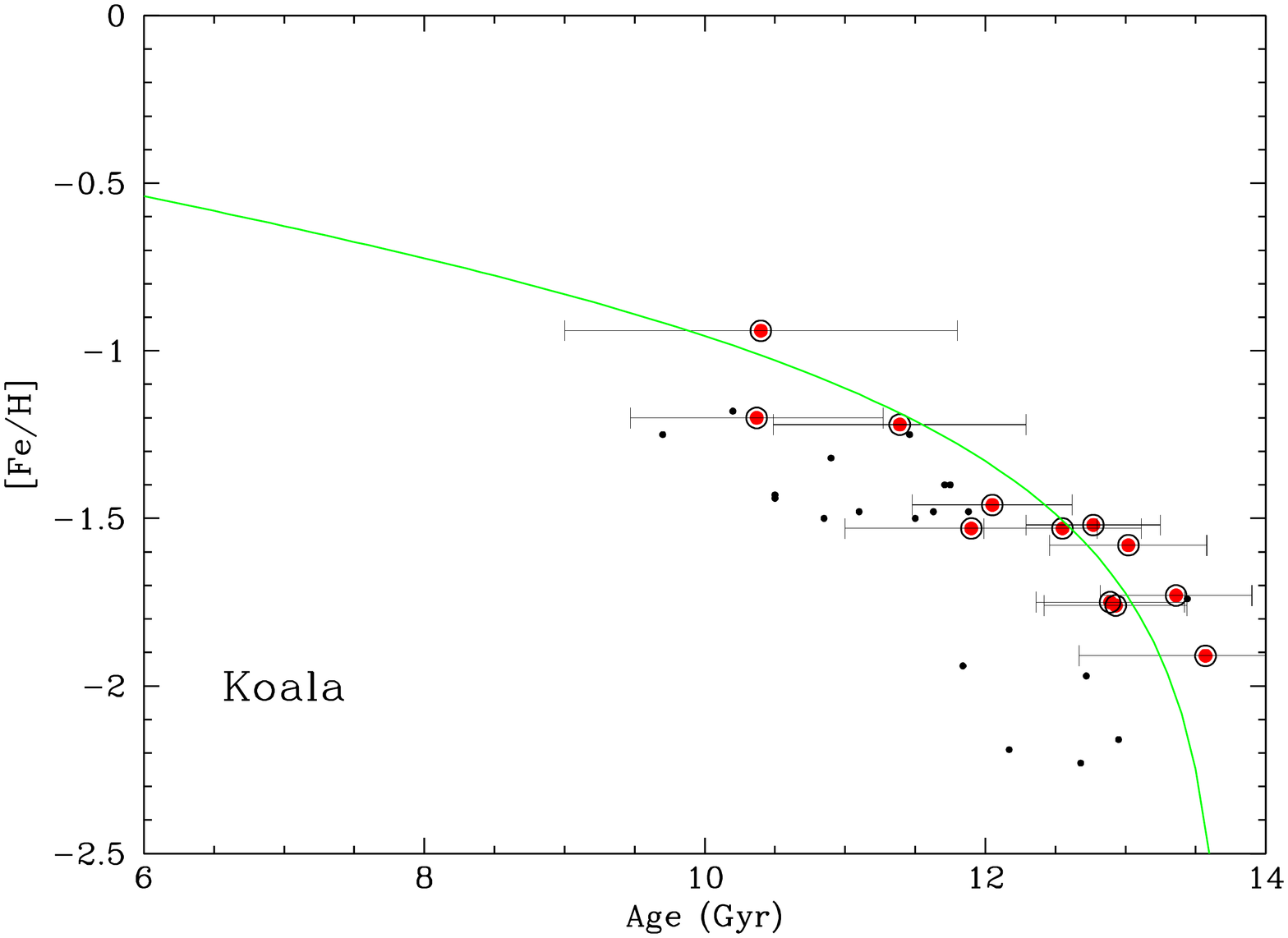}
    \caption{Koala AMR. Red symbols show GC assigned to Koala. Small black symbols show the remaining GCs of the MW. The green line is the best-fit AMR to the GCs. }
    \label{koala}
\end{figure}

Of the 21 probable GCs, a dozen have available ages and metallicities which we show in Fig.~\ref{koala}. Due to the large scatter in the data points, we were unable to obtain a well-defined fit to the Koala AMR with both the yield and age of formation as free parameters. For this system, we fix the age to that of the Universe (i.e. 13.72 Gyr in our cosmology) and derive an effective yield of 
$p$ = 0.35$\pm$0.05. In this case the measurement uncertainty does not capture the possible systematic uncertainty due to using a fixed formation age. 
The high measured yield suggests that Koala was a fairly massive satellite.


\subsection{Helmi Streams}

Based on their IOM from Gaia, Koppelman et al. (2019) assigned 7 GCs to the Helmi streams. Two of these (NGC 5634 and NGC 5904) M19 list as tentative due to a possible association with Gaia-Enceladus. M19 also include Rup106 and Pal5 as tentative Helmi streams progenitor GCs. 

\begin{figure}
	\includegraphics[width=0.8\columnwidth,angle=-90]{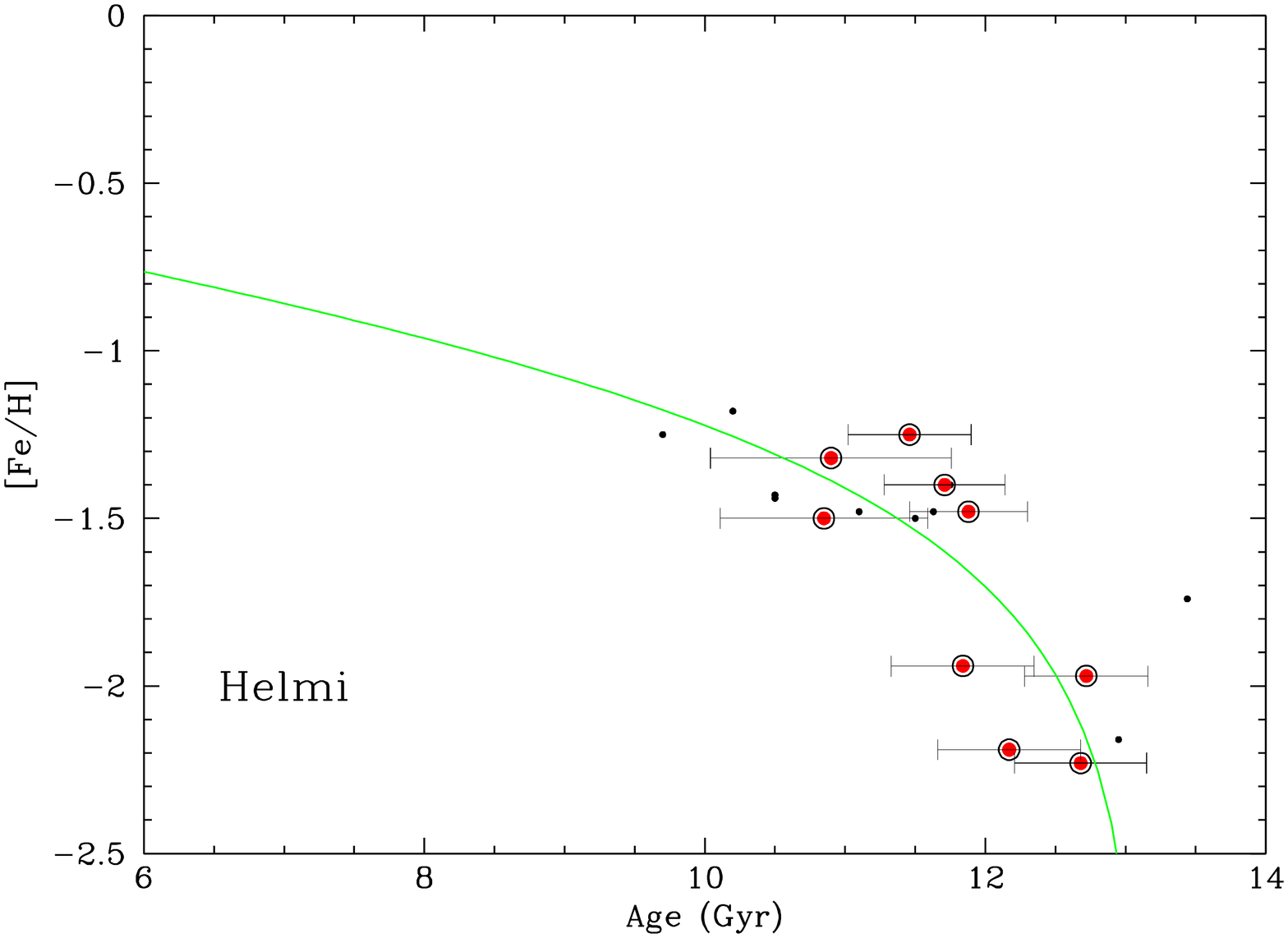}
    \caption{Helmi streams AMR. Red symbols show GCs assigned to the Helmi streams. 
    Small black symbols show the remaining GCs of the MW. The green line is the best-fit AMR to the GCs. }
    \label{helmi}
\end{figure}

In Fig.~\ref{helmi} we show these 5 probable GCs and 4 tentative GCs associated with the Helmi streams. We find that all 9 of them are well represented by an AMR with $p$ = 0.22$\pm$0.06, and we assign all 9 to be probable GCs of the Helmi streams progenitor. 


The N-body model of Koppelman et al. (2019) assigns a stellar mass of $\sim$10$^8$ M$_{\odot}$ and halo mass of $\sim$10$^{10}$ M$_{\odot}$ 
to the Helmi streams progenitor and suggests it was accreted 5--8 Gyr ago. 

\subsection{High-Energy Group}

\begin{figure}
	\includegraphics[width=0.8\columnwidth,angle=-90]{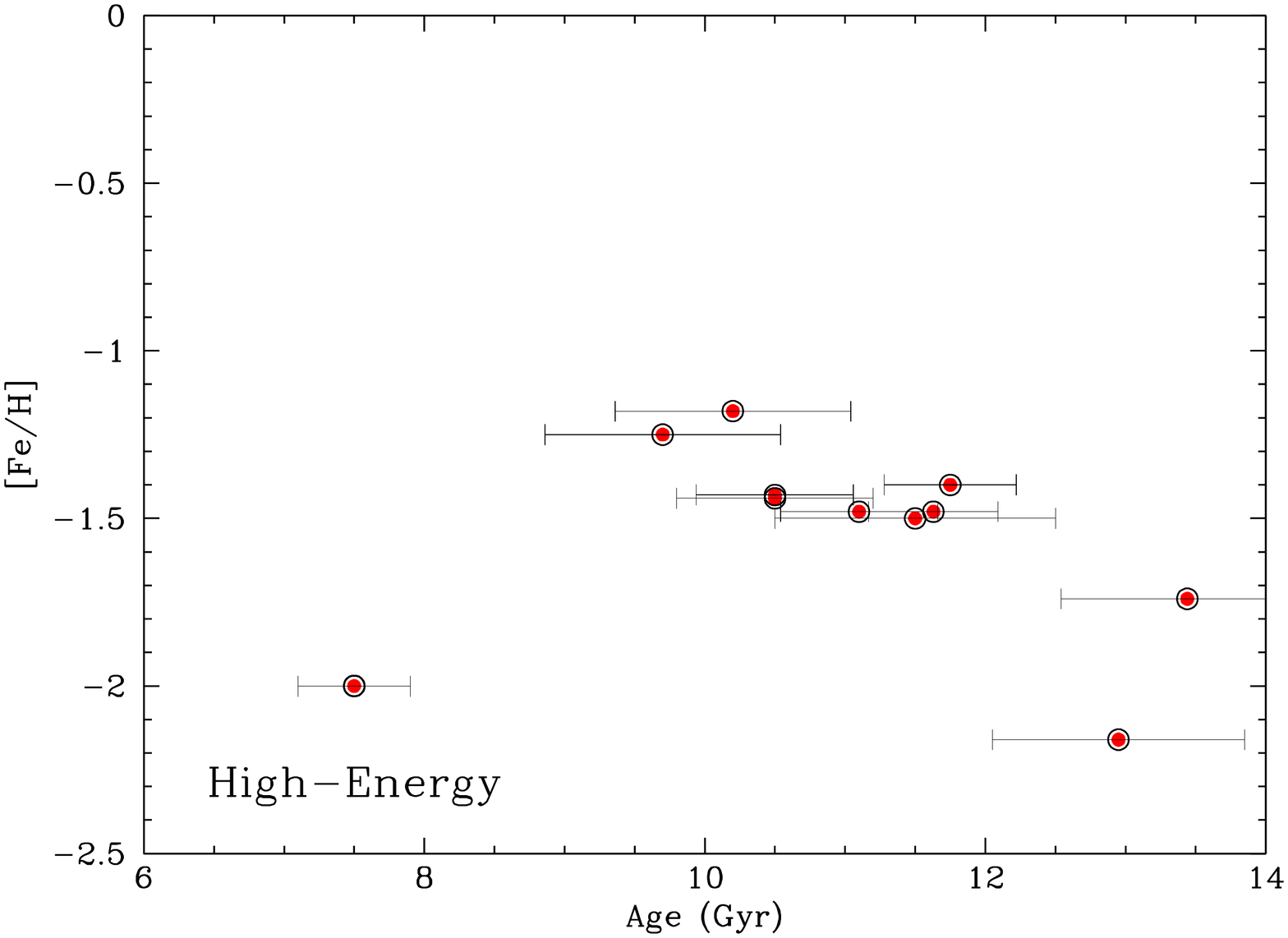}
    \caption{High-energy group. Red symbols show high-energy GCs. The GC at 7.5 Gyr and [Fe/H] = --2 is Crater (see text for details).  }
    \label{he}
\end{figure}

M19 find 11 GCs to have high energies but a large range in angular momenta, suggesting they came from different progenitors. Thus they are not expected to follow a single AMR. Ten of these have ages and metallicities available from K19, which we show in Fig.~\ref{he}. The majority of these GCs would appear to have been accreted from  progenitor(s) of low mass similar to the Helmi streams progenitor. 

Although the high-energy GC Crater does not have an age and metallicity from K19, it has been observed using HST by Weisz et al. (2016). They find a good fit to the CMD using a single stellar population model with [$\alpha$/Fe] = +0.4, age = 7.5 $\pm$ 0.4 Gyr, [M/H] = --1.66 $\pm$ 0.04 (corresponding to [Fe/H] $\sim$ --2). Plotting it on Fig.~\ref{he} suggests Crater came from a very low mass progenitor (a conclusion also favoured by Weisz et al. 2016).

\section{Derived Properties}

In Table 1 we summarise the probable GCs assigned to each progenitor satellite, including 1 tentative GC (Pal~2) assigned to Gaia-Enceladus (with details in Appendix A). These assignments differ somewhat from those given in M19 largely due to our increased sample of ages and metallicities plus the use of alpha-element ratios. We also list the GC that is likely to be the former nucleus of the progenitor satellite for Sgr, Gaia-Enceladus and Sequoia. The other satellites may have had nuclear star clusters but it is unclear which GC, if any, was its former nucleus.  A total of 76 GCs 
are assigned to 5 progenitor satellites. 

Burkert \& Forbes (2019) showed that the number of GCs in a galaxy is an excellent predictor of the total halo mass even for low mass galaxies with a handful of GCs. Boylan-Kolchin (2017) and Burkert \& Forbes (2019) argued that the relation is set at very early times with GCs forming within galactic-sized dark matter halos, and that subsequent mergers largely move galaxies parallel to the (linear) relation. 

Combining the 
probable and tentative GCs we estimate the total number of GCs from each progenitor galaxy. 
Some GCs can clearly survive a merger with the MW (during which its host galaxy is completely disrupted). It is difficult to know, however, if any GCs have been destroyed in that process. 
We include in our calculation the possible nuclear cluster (which is now classified as a Milky Way GC), which may lead to an overestimate of the original GC count (complicated by the fact that the nucleus of dwarf galaxies may be the result of in-spiralling GCs). With these caveats in mind, we use eq. 1 of Burkert \& Forbes (2019) to  obtain a halo mass of the satellite, under the assumption of a $\Lambda$CDM cosmology. 


From this halo mass we estimate the stellar mass using the standard zero redshift stellar mass--halo mass (SMHM) relation. 
In particular we use eq. 66 and table 6 of 
Rodriguez-Puebla et al. (2017), after noting the typo in the equation with +1/2 should be -1/2, to estimate the total stellar mass of the progenitor satellite. Here we use the parameters for z = 0.1, but note that the evolution of the SMHM is minor for low-mass galaxies. For example, the difference in the log halo mass for a galaxy with stellar mass of $\sim$10$^8$ M$_{\odot}$ at z = 0.1 and z = 2 (10.4 Gyr ago) is less than 0.05 dex. Thus the predicted stellar mass is not particularly sensitive to the epoch in which the satellite merged with the MW.

A well-known relation also exists between the total stellar mass and the mean metallicity of a galaxy today. 
We use 
eq. 4 of Kirby et al. (2013) for Local Group dwarf galaxies to estimate the mean [Fe/H] metallicity of the progenitor satellite. Given that the normalisation of the mass-metallicity relation varies with redshift, this inferred metallicity is an upper limit to the metallicity of the satellite at infall. In the study of Mannucci et al. (2009) the metallicity at z $\sim$ 2 (10.4 Gyr ago) is about 0.8 dex lower than today for a 10$^8$ M$_{\odot}$ galaxy. 

Table 2 lists the approximate halo mass, stellar mass and upper limit to the mean metallicity for each satellite at the time of infall onto the MW. 
Table 2 also lists the results of our AMR fitting, i.e. the effective yield and the age when the system first formed according to eq. 1. For Koala we list the fixed age assumed in the AMR fit.

\begin{table}
	\centering
	\caption{Accreted Satellites: Globular Clusters}
	\label{tab:table1}
	\begin{tabular}{lcc} 
		\hline
		Progenitor & GCs & Nucleus \\
		Galaxy & & \\
		\hline
		Sgr & 9 & M54 \\
		Gaia-E & 27+1 & NGC~1851 \\
				Sequoia & 9 & $\omega$-Cen \\
						Koala & 21 & -- \\
								Helmi & 9 & -- \\
					High-E & 11 & -- \\
		\hline
	\end{tabular}
\end{table}

\begin{table}
	\centering
	\caption{Accreted Satellites: Derived Properties}
	\label{tab:table2}
	\begin{tabular}{lcccccr} 
		\hline
		Progenitor & logM$_{h}$ & logM$_{\ast}$ & [Fe/H] & Yield & Age\\
		Galaxy & (M$_{\odot}$) & (M$_{\odot}$) & (dex)  & & (Gyr)\\
		\hline
		Sgr & 10.64 & 7.9 & $<$-1.1 & 0.33$\pm$0.06 & 13.24$\pm$0.19\\
		Gaia-E & 11.14  & 8.9 & $<$-0.8 & 0.27$\pm$0.02 & 13.55$\pm$0.10\\
		Sequoia & 10.64 & 7.9 & $<$-1.1 & 0.26$\pm$0.03 & 13.34$\pm$0.11 \\
		Koala & 11.01 & 8.7 & $<$-0.9 & 0.35$\pm$0.05 & 13.72\\
		Helmi & 10.64 & 7.9 & $<$-1.1 & 0.22$\pm$0.06 & 13.13$\pm$0.68\\
		\hline
	\end{tabular}
\end{table}

\begin{table}
	\centering
	\caption{Accreted Satellites: Literature Properties}
	\label{tab:table3}
	\begin{tabular}{lcccr} 
		\hline
		Progenitor & logM$_h$ & logM$_{\ast}$ & Infall\\
		Galaxy & (M$_{\odot}$) & (M$_{\odot}$) & (Gyr)\\
		\hline
		Sgr & $>$10.8 & 7.99-8.16 & 8-9\\
		Gaia-E & 11-11.7 & 8.77-9.30 & 9-11\\
		Sequoia & $\sim$10 & $\sim$7.70 & $\sim$10\\
		Koala & -- & -- & --\\
		Helmi & $\sim$10 & $\sim$8.0 & 5-8\\
		\hline
	\end{tabular}
\end{table}

\begin{figure}
	\includegraphics[width=0.8\columnwidth,angle=-90]{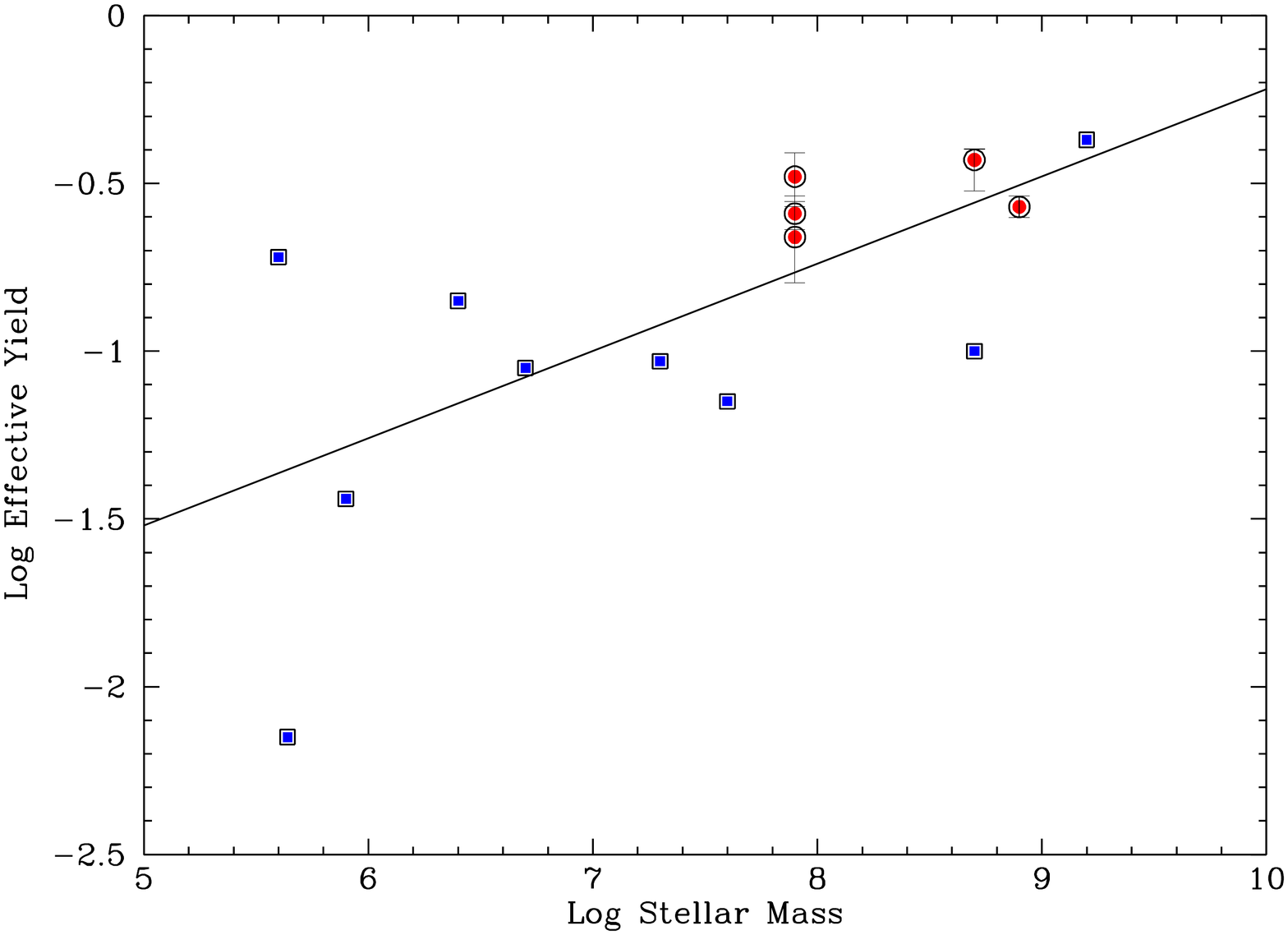}
    \caption{Effective yield vs stellar mass. The effective yield is from a leaky-box chemical evolution model fit. Red circle symbols show our results from fits to the GCs of the five disrupted satellites (see Table 2). Blue square symbols show values from Leaman et al. (2013b) from fits to the stellar component of Local Group galaxies (errors are on the order of the symbol size). The solid line shows the best fit of slope 0.28, a similar slope is achieved for galaxies with log M$_{\ast}$ $>$ 7.    }
    \label{p}
\end{figure}

Given that the mean metallicity of a system is driven by the effective yield $p$ (which we determine by a  leaky-box chemical evolution model fit to the GC ages and metallicities) it should be related to the stellar mass of the host galaxy. In Fig.~\ref{p} we show our measured effective yields vs the stellar mass we derive above. We also include several Local Group galaxies (i.e. LMC, SMC, WLM, Fornax, LeoI, LeoII, Carina, Sculptor and Sextans) using results from Leaman et al. (2013b). They fit a similar leaky-box model to the field stars of these galaxies. We take their stellar masses from the work of McConnachie (2012). 

A best fit correlation is also show on Fig.~\ref{p}. It has the form:\\

log $p$ = 0.26($\pm$0.08) logM$_{\ast}$ -- 2.82($\pm$0.55)\\

We find a similar slope if we restrict the sample to galaxies with logM$_{\ast}$ $>$ 7. The slope we find is in excellent agreement with the Kirby et al. (2013) relation slope (i.e. 0.30$\pm$0.02) between mean metallicity and stellar mass for Local Group dwarf galaxies. 
As our yields and stellar masses are derived from independent quantities, the presence of a correlation and its slope give us confidence in our methodology. 

In Table 3 we list estimates of the satellite mass and infall epoch for several accreted satellites that were we able to find in the literature. Our estimates for the halo and stellar masses 
(see Table 2) are in reasonable agreement with the literature estimates. Note that the literature estimates from Myeong et al. (2019) for Sequoia and Gaia-Enceladus are based on the mass of their GC systems.



\section{Discussion}

In this work we have taken advantage of the 
new [Si/Fe] abundances from SDSS/APOGEE survey for 46 GCs published by Horta et al. (2020). Using this information and IOMs they examined the in-situ and accretion origins for their sample of GCs. They discuss several individual GCs. As mentioned in section 4 we follow Horta et al. by assigning three GCs (i.e. NGC 6121, NGC 6441 and Pal~6) to be in-situ, rather than accreted as assigned by M19. Horta et al. also mention a few other GCs and we agree with them concerning Pal 10 (in-situ), NGC 288 (accreted), NGC 5904 (accreted). They suggest a `possible accreted' origin for NGC 6388 -- here we continue to place it firmly in the in-situ category (as does M19) based on its location in age-metallicity and IOM space.

From an initial sample of 160 MW GCs, we assign 
73 to have formed in-situ within the early MW. It is likely that 8 GCs with neither Gaia kinematics nor ages and metallicities were also formed in-situ.  
We assign 76 GCs to 5 progenitor satellites. The remaining 11 accreted GCs are likely associated with low-mass (M$_{\ast}$ $<$ 10$^8$ M$_{\odot}$) satellites. The 5 identified satellites have estimated stellar masses of 10$^{8-9}$ M$_{\odot}$, halo masses of  10$^{10.6-11.1}$ M$_{\odot}$ and metallicities --1.1 $<$ [Fe/H] $<$ --0.8. 
At least 3 of them were likely nucleated dwarf galaxies. From literature studies, it is suggested they were accreted at similar times, between 8 and 11 Gyr ago (except perhaps for the Helmi streams satellite which may have had its infall as recently as 5 Gyr ago). This may indicate that the satellites were accreted as part of a group of satellites.  
We note that the spatial distribution of existing Milky Way satellites is not isotropic but rather they form a planar disk (Pawlowski et al. 2014).  
Initial star formation in each satellite appears to have occurred over 13 Gyr ago and the oldest GCs formed well before the infall of the satellite.

Each of the five satellites are dominated by metal-poor GCs, with Sgr being a notable exception having 4 relatively metal-rich ([Fe/H] $>$ --1) GCs. Both Sgr and Gaia-Enceladus may have associated open clusters of age $\sim$4--7 Gyr. If correct, this indicates their formation occurred  after infall and that the progenitor satellites were gas-rich. 
The lack of any satellites with masses comparable to the MW is consistent with the limits from Galactic disk stars suggesting no major disruptive mergers in the last 7--10 Gyr (Burnett et al. 2011). Nevertheless, our results point to the growth of the MW's GC system being dominated by a small number of relatively massive satellites.

The MW's halo can be described as a `dual halo' with an inner halo 
of mean metallicity [Fe/H] $\sim$ --1.6 and an outer halo of [Fe/H] $\sim$ --2.2 (Carollo et al. 2007). Although both components are expected to have contributions from accreted stars, it is the outer halo that is dominated by accreted stars (Tissera et al. 2011). In the suite of 8 MW-like model galaxies in the simulation by Tissera et al. (2011) the stars accreted from satellites have  ages between 9.8 and 12.6 Gyr and metallicities of --1.45 $<$ [Fe/H] $<$ --0.86.  
Such ages and metallicities are consistent with the initial epoch of star formation and metallicity upper limits for the five disrupted satellites in our work (see Table 2). Using MW-like model galaxies based on the EAGLE simulations by Mackereth et al. (2019) concluded that the MW's assembly history was dominated by the accretion of a small number of 
satellites of stellar mass 10$^{8.5-9}$ M$_{\odot}$, in reasonable agreement with our findings (see Table 2).


\subsection{The Kraken/Koala satellite}

K19 describes their simulation of 25 galaxies with MW-like masses within the EAGLE galaxy formation model (Crain et al. 2015). Their suite of MW-like model galaxies undergoes an average of 4.4$\pm$2.3 minor mergers.  with mass ratios of 1:100 up to 1:4. 
This corresponds to satellite masses of 10$^{8-9}$ M$_{\odot}$ at the time of infall. They suggest that the vast bulk of the accreted GCs come from two nearly equal mass ($\sim$10$^{9}$ M$_{\odot}$) 
satellites than merged in the last 10 Gyrs. A third lower-mass ($\sim$10$^{8}$ M$_{\odot}$) 
satellite brings in a handful of GCs. Additional accreted satellites are of very low mass, containing few, if any, GCs. 

They associated the low-mass satellite with the Sgr dwarf galaxy and one of the massive satellites with the Canis Major dwarf galaxy (here called Gaia-Enceladus). The remaining, and perhaps most massive satellite, was not associated with any known feature of the MW by K19 but instead was given the name Kraken. They suggested the Kraken had a stellar mass of 
2 $\times$ 10$^{9}$ M$_{\odot}$ and was accreted 6--9 Gyr ago. Although K19 do not 
associate any particular MW GC with the Kraken, they do 
list 18 GCs with the potential to be from the disrupted Kraken satellite. 

Here we have assigned those 18 GCs across the Gaia-Enceladus, Sequoia, Koala, Helmi satellites and the high-energy group. So, while we do suggest that the MW accreted an additional satellite of mass M$_{\ast}$ $\sim$  10$^{9}$ M$_{\odot}$ (i.e. Koala), we make different GC assignments to K19. In their terminology, satellite 3 is the Sgr dwarf, satellite 2 is 
Gaia-Enceladus/Canis Major and satellite 1 could be the Koala galaxy with $\omega$-Cen as its former nucleus. Our analysis indicates that the MW has accreted 5 satellites of mass 10$^{8-9}$ M$_{\odot}$ consistent with K19.

Overall, K19 concluded that the MW formed quite rapidly with no major mergers since z = 4 (i.e. the last 12.3 Gyr). This makes it somewhat atypical compared to their suite of 25 MW-like models. 

\section{Conclusions}

Using the integrals of motion (IOM), the age-metallicity relations (AMRs) and alpha-element ratios of Milky Way globular clusters (GCs) we have assigned a total of 76 GCs to 5 progenitor satellite galaxies that have been accreted and subsequently disrupted. Although the satellites have been disrupted their GCs reveal long-lasting signatures in their IOM, AMRs and alpha-element ratios. 

From the number of GCs we use various scaling relations to estimate their original halo mass, stellar mass and upper limit on their mean metallicity. We find a reasonable correlation between the inferred stellar mass and  the effective yield from fitting a leaky-box chemical enrichment model to the AMR of their GCs. 
Based on these results, we reverse engineer the Milky Way to identify some characteristics of these 5 accreted satellites. \\

\noindent
{\bf Sagittarius}\\
This well-known accreted satellite was likely a gas-rich nucleated dwarf galaxy with 9 known GCs (the GC M54 being its former nucleus). It had an original stellar mass of log M$_{\ast}$ $\sim$ 8 and was accreted 8--9 Gyr ago.\\

\noindent
{\bf Gaia-Enceladus}\\
A gas-rich nucleated dwarf galaxy of stellar mass log M$_{\ast}$ $\sim$ 9, making it the most massive identified satellite to be accreted by the Milky Way. We suggest that the GC NGC 1851 was its former nucleus. It hosted over two dozen GCs. Many of its GCs have previously been associated with the Canis Major dwarf galaxy. It reveals a particularly well-defined age-metallicity relation and was accreted 9--11 Gyr ago. \\

\noindent
{\bf Sequoia}\\
A nucleated dwarf galaxy of stellar mass log M$_{\ast}$ $\sim$ 8. We suggest that the GC $\omega$-Cen is its former nucleus. It was accreted some 10 Gyr ago.\\  

\noindent
{\bf Koala}\\
A low-energy satellite with 21 associated GCs which we dub Koala. It had an original stellar mass of log M$_{\ast}$ $\sim$ 8.7. Koala might be the equivalent of satellite 1, called the Kraken, in Kruijssen et al. (2019). \\

\noindent
{\bf Helmi streams}\\
A low mass satellite of original stellar mass
log M$_{\ast}$ $\sim$ 8 accreted between 5 and 8 Gyr ago. It contained at least 9 GCs. \\

 The 5 identified satellites have a total stellar mass of log M$_{\ast}$ $\sim$ 9 and hence only represent a few percent of the Milky Way's current stellar mass. However, their 76 GCs represent a significant fraction of the Milky Way's GC system. 
A small number of GCs (11) are consistent with being accreted but can not be assigned to a single progenitor satellite.
These GCs likely came from a number of low mass (log M$_{\ast}$ $\le$ 8) satellites.
We estimate that, in total, more than half of the Milky Way's GC system has been accreted from a number of disrupted satellites.

When more alpha-element ratios are available for MW GCs a multi-dimensional statistical analysis could be applied to gain further insight. 
 Furthermore, a similar analysis to that carried out here could be applied to the M31 GC system and its distinct kinematic groups (Mackey et al. 2019).

\section*{Acknowledgements}

We thank the ARC for financial support via DP160101608. We thank A. Alabi, J. Brodie and J. Pfeffer for useful discussions. We thank the referee for their useful comments.




\bibliographystyle{mnras}
\bibliography{MWGCs} 




\appendix

\section{List of Accreted Globular Clusters} 

\subsection{Sagittarius (9 GCs)}

Probable membership: NGC 2419, NGC 5824, NGC 6715 (M54), Arp 2, Pal 12, Terzan 7, Terzan 8, Whiting 1 and AM4. 

\subsection{Gaia-Enceladus (28 GCs)}

Probable membership: NGC 288, NGC 362, NGC 1261, NGC 1851, NGC 1904, NGC 2298, NGC 2808, NGC 4147, NGC 4833, NGC 5286, NGC 5897, NGC 6205, NGC 6229, NGC 6235, NGC 6284, NGC 6341, NGC 6779, 
NGC 6864, NGC 7089, NGC 7099, NGC 7492, 
IC 1257, Djorg 1, Terzan 10, ESO-SC06, Pal 1 and Pal 15. \\ 
Tentative membership: Pal 2. 

\subsection{Sequoia (9 GCs)}

Probable membership: NGC 3201, $\omega$-Cen (NGC 5139), NGC 5466, NGC 6101, NGC 6535, NGC 7006, IC 4499, FSR1758 and Pal 13.

\subsection{Koala (21 GCs)}

Probable membership: NGC 5946, NGC 5986, NGC 6093, 
NGC 6139, NGC 6144, NGC 6254, NGC 6256, NGC 6273, NGC 6287, NGC 6333, NGC 6401, 
NGC 6402, NGC 6453, NGC 6517, 
NGC 6541,  NGC 6544, NGC 6681, NGC 6712 and NGC 6809, FSR1735  
and Ton 2.

\subsection{Helmi Streams (9 GCs)}

Probable membership: NGC 4590, NGC 5024, NGC 5053, NGC 5272, NGC 5634, NGC 5904, NGC 6981, Pal 5 and Rup 106.  

\subsection{High-Energy Group (11 GCs)}

Probable membership: NGC 5694, NGC 6426, NGC 6584, NGC 6934, AM1, Eridanus, Pyxis, Pal 3, Pal 4, Pal 14 and Crater.




\bsp	
\label{lastpage}
\end{document}